\documentclass[aps,prd,preprintnumbers,superscriptaddress,nofootinbib
]{revtex4}%
\usepackage[dvipdfmx]{graphicx} 
\usepackage{feynmf}
\usepackage{bm,latexsym,amsmath,amssymb,amsfonts,mathrsfs}
\usepackage[]{hyperref}

\usepackage{physics} 
\usepackage{tensor}
\usepackage{ulem}

\usepackage{physics}
\usepackage{aas_macros}

\newcommand*{\beqa}{\begin{eqnarray}}
\newcommand*{\eeqa}{\end{eqnarray}}

\newcommand*{\p}{\partial}

\newcommand{\x}{{\boldsymbol{x}}}

\newcommand{\bphi}{{\boldsymbol{\phi}}}

\begin{document}

\title{Hidden Symmetries of Power-law Inflation}

\author{
Takeshi Chiba
}
\affiliation{Department of Physics, College of Humanities and Sciences, Nihon University, \\
Tokyo 156-8550, Japan}
\author{
Tsuyoshi Houri
}
\affiliation{National Institute of Technology, Maizuru College, \\
Kyoto 625-8511, Japan
}
\date{\today}

\begin{abstract}
A scalar field with an exponential potential 
has been proposed as a model of inflation (called power-law inflation).
Although it admits an exact solution, the integrability of the system has
not been shown.
We uncover the hidden symmetries behind the system
by utilising the Eisenhart lift of field theories.
We find that a conformal Killing vector field in the field space
exists only for a particular combination of exponential functions
that includes a single exponential potential.
This implies the existence of
additional conserved quantity and explains the integrability of the system.
\end{abstract}


\maketitle

\section{Introduction}

Power-law inflation \cite{Lucchin:1984yf} is a model of inflation in
which the scale factor grows as a power-law of cosmic time instead of exponential of time. It is described by a scalar field with an exponential
potential of the form $V(\phi)=V_0e^{-\lambda\phi}$ (in the reduced Planck units $8\pi G=1$) \cite{2000cils.book.....L}.
A remarkable property of the model is that
it admits exact solutions of equations of motion of the scalar
factor and the scalar field and hence is integrable.
This fact suggests the existence of symmetries of the system.
However, such symmetries for general $\lambda$ have not been found
so far.

An important step toward revealing the
hidden symmetries of power-law inflation was taken
in \cite{deRitis:1990ba} where the Noether symmetry
for the Lagrangian was studied. It was found there that such a symmetry exists
only for $\lambda=\sqrt{6}/2$. However, to our knowledge, no
symmetries for general $\lambda$ have been found.
Although power-law inflation is ruled out as a model of inflation by
the Planck data \cite{Planck:2013jfk}, exponential potentials appear
ubiquitously from the perspective of
higher dimensional theories such as string theory and/or by changing
the frame (Jordan vs. Einstein), and the study of symmetries of such a system may be interesting in its own right.

In this paper, instead of finding symmetries of the Lagrangian,
we search for symmetries of the system of a scalar field in
Friedmann-Lemaitre-Robertson-Walker (FLRW) universe by applying
the Eisenhart lift of scalar field theory introduced in \cite{Finn:2018cfs}
to the system.
The equations of motion of the scale factor and the scalar field then
become geodesic equations for null geodesics in lifted field space.
Thus, finding the conserved quantities of the system is reduced to finding the conserved quantities along null geodesics.
We find that a nontrivial conformal Killing vector field exists
for a particular combination of exponential functions which includes a single exponential potential with general $\lambda$:
the ``hidden symmetry'' of power-law inflation is revealed.\footnote{
We also find that another conformal Killing vector field exists for general $V(\phi)$. See footnote in Sec.\ref{sec3B}.}

The paper is organized as follows.
After reviewing Eisenhart lift for
scalar fields in Sec.~\ref{sec2},
the system of a homogeneous scalar field
in FLRW universe is lifted, and the existence of a conformal Killing
vector field for a particular combination of exponential functions
and the relation to the previous work are discussed in Sec.~\ref{sec3}.
Sec.~\ref{sec7} is devoted to summary.
In Appendix \ref{appendixA}, the details of solving the conformal Killing equations are described.

Our convention of the metric signature is $(-,+,+,+)$ and we use
the units of $8\pi G=c=1$.

\section{Eisenhart Lift for Scalar Fields}
\label{sec2}

After reviewing the Eisenhart lift for a particle, the Eisenhart
lift for scalar fields is introduced.

\subsection{Eisenhart Lift for a Particle}

Consider the equation of motion of a particle with unit mass
in a three-dimensional space
\beqa
\ddot x^{i}=-\p_{i}V\,,
\label{eom1}
\eeqa
where $i=1,\dots ,3$ and $\ddot x^i=d^2x^i/dt^2$ with $t$
being the time coordinate and $\p_iV=\p V/\p x^i$.
The equation of motion is derived from the action
\beqa
S_p=\int dt \left(\frac12 \dot \x^2-V(\x)\right)\,.
\label{action1}
\eeqa
Eisenhart showed that the same dynamics can be derived from the following action by adding a new coordinate $y$ \cite{eisenhart1928} (see \cite{Cariglia:2014ysa,Cariglia:2015bla} for a review)
\beqa
S'_p=\int dt\left(\frac12 \dot\x^2+\frac{1}{4V(\x)}\dot y^2\right)\,.
\label{action2}
\eeqa
In fact, the equation of $\x$ and $y$ are
\beqa
\ddot x^i&=&-\frac{1}{4V^2}\dot y^2\p_iV\,,
\label{eom2-1}\\
p_y&=&\frac{\dot y}{2V}={\rm const}\,.
\label{eom2-2}
\eeqa
So, plugging Eq. (\ref{eom2-2}) into Eq. (\ref{eom2-1}) and
setting $p_y=1$ reproduces Eq. (\ref{eom1}). Changing
$p_y$ merely corresponds to the rescaling of the time coordinate.

Introducing  $x^A=(x^i,y)$,
the action Eq.(\ref{action2}) may be viewed as that of a free particle in four-dimensional space:
\beqa
S_p'=\int dt~ \frac12 G_{AB}\dot x^A\dot x^B
\eeqa
where $G_{AB}$ is the field space metric with
$G_{ij}=\delta_{ij}, G_{iy}=0$ and $G_{yy}=\frac{1}{2V}$.
Hence, finding the conserved quantities of
the system is reduced to finding the conserved quantities for
geodesics.

\subsection{Eisenhart Lift for Scalar Fields }
\label{sec2B}

The Eisenhart lift is extended to scalar field theories by \cite{Finn:2018cfs}.
Consider the system of $n$ scalar fields in a four-dimensional spacetime
(with the metric $g_{\mu\nu}$)
\beqa
S=\int d^4x\sqrt{-g}\left(\frac12 R-\frac12 g^{\mu\nu}k_{IJ}(\bphi)\p_{\mu}\phi^I\p_{\nu}\phi^J
-V(\bphi)\right)\,,
\label{action:scalar}
\eeqa
where the first term is the Einstein -Hilbert term (in units of $8\pi G=1$)
and $I=1,\dots,n$ is the field space index and $k_{IJ}(\bphi)$ is
the scalar field space metric, while $\mu$ is the spacetime index.
The Einstein equation and the equation of motion of $\phi^I$ are given by
\beqa
&& R_{\mu\nu}-\frac12 g_{\mu\nu}R=k_{IJ}\p_{\mu}\phi^I\p_{\nu}\phi^J-\frac12 g_{\mu\nu}\left(g^{\alpha\beta}k_{IJ}\p_{\alpha}\phi^I\p_{\beta}\phi^J
+2V \right)
\label{einstein:scalar1}\\
&& \Box\phi^I+\Gamma^{I}_{JK}g^{\mu\nu}\p_{\mu}\phi^J\p_{\nu}\phi^K-k^{IJ}\p_JV=0\,,
\label{eom:scalar1}
\eeqa
where $\Gamma^I_{JK}$ is the Christoffel symbol constructed from the
field space metric $k_{IJ}$.

\cite{Finn:2018cfs} observed that the same dynamics can be described by the
following Eisenhart lift through the introduction of the
fictitious vector field $B^{\mu}$
\beqa
I_R=\int d^4x\sqrt{-g}\left(\frac12 R-\frac12 g^{\mu\nu}k_{IJ}(\bphi)\p_{\mu}\phi^I\p_{\nu}\phi^J
+\frac{1}{4V(\bphi)}\left(\nabla_{\mu}B^{\mu}\right)^2
\right)\,.
\label{action:scalar2}
\eeqa
The Einstein equation and the equations of motion of $\phi^I$ and $B^{\mu}$ are
\beqa
&&R_{\mu\nu}-\frac12 g_{\mu\nu}R=k_{IJ}\p_{\mu}\phi^I\p_{\nu}\phi^J-
\frac12 g_{\mu\nu}\left(g^{\alpha\beta}k_{IJ}\p_{\alpha}\phi^I\p_{\beta}\phi^J\right)\nonumber\\
&&~~~~~~~~~~~~~~~~~~~~+2B_{(\mu}\p_{\nu)}\pi_B-g_{\mu\nu}B^{\alpha}\p_{\alpha}\pi_B-g_{\mu\nu}V\pi_B^2
\label{einstein:scalar2}
\\
&&\Box\phi^I+\Gamma^{I}_{JK}g^{\mu\nu}\p_{\mu}\phi^J\p_{\nu}\phi^K-
\pi_B^2k^{IJ}\p_JV=0\,,
\label{eom:scalar2-1}
\\
&&\p_{\mu}\pi_B=0\,,
\label{eom:scalar2-2}
\eeqa
where
\beqa
\pi_B=\frac{\nabla_{\mu}B^{\mu}}{2V}\,.
\label{piB}
\eeqa
{}From Eq. (\ref{eom:scalar2-2}), $\pi_B$ is a constant.
Plugging this into Eq. (\ref{einstein:scalar2}) and Eq. (\ref{eom:scalar2-1}) and setting $\pi_B=1$ reproduces Eq. (\ref{einstein:scalar1}) and Eq. (\ref{eom:scalar1}).

\section{Integrable Cosmology: Scalar Field in FLRW Universe}
\label{sec3}

We apply the formalism of the Eisenhart lift of scalar fields to
the system of a single scalar field in Friedmann-Lemaitre-Robertson-Walker
(FLRW) universe \cite{Cariglia:2018mos}.

\subsection{Field Space Metric and the Equations of Motion}
\label{sec3A}

We consider a flat FLRW universe $g_{\mu\nu}dx^{\mu}dx^{\nu}=-N(t)^2dt^2+a(t)^2d\x^2$ where $N(t)$ is the lapse function and $a(t)$ is the scale factor.
Assuming that a scalar field $\phi$ and a vector field
$B^{\mu}$ are homogeneous,
the lifted system Eq. (\ref{action:scalar2}) reduces to the particle system
whose Lagrangian is
\beqa
{\cal L}=-\frac{3a}{N}\dot a^2+\frac{a^3}{2N}\dot\phi^2+\frac{1}{4Na^3V}\dot\chi^2
\equiv \frac12 G_{AB}\dot\phi^A\dot\phi^B\,,
\eeqa
where $\chi\equiv Na^3B^0$, $\phi^A=(a,\phi,\chi)$, and the field space metric $G_{AB}$ is given by
\beqa
G_{AB}=
\begin{pmatrix}
-\frac{6a}{N} & & \\
& \frac{a^3}{N}& \\
& &\frac{1}{2Na^3V} \\
\end{pmatrix}
\label{effective-metric}
\,.
\eeqa
In terms of the conjugate momenta, $p_a=-6a\dot a/N,
p_{\phi}=a^3\dot\phi/N, p_{\chi}=\frac{\dot\chi}{2Na^3V}$ which coincides with $\pi_B$ in Eq. (\ref{piB}), the Hamiltonian becomes
\beqa
{\cal H}=\frac12G^{AB}p_Ap_B=N\left(-\frac{1}{12a}p_a^2+\frac{1}{2a^3}p_{\phi}^2+a^3V(\phi)p_{\chi}^2\right)\,,
\eeqa
and variation of ${\cal H}$ with respect to $N$ yields the Hamiltonian constraint
\beqa
H= \frac{1}{2}\left(
-\frac{p_a^2}{6a}+\frac{p_\phi^2}{a^3}
+2a^3V(\phi)p_\chi^2 \right)=0\,.
\label{H-constraint}
\eeqa
Therefore, $p_A$ is a null vector.
Henceforth, we set $N=1$.

The equations of motion of $\phi^A$ are null geodesic equations of $G_{AB}$
and are derived from
the canonical equations of motion, $\dot\phi^A=\frac{\p H}{\p p_A}, \dot p_A=-\frac{\p H}{\p \phi^A}$. Along with the Hamiltonian constraint Eq. \eqref{H-constraint}, they are given by
\beqa
&& \frac{\ddot a}{a}=-\frac{1}{3}\left(\dot\phi^2-Vp_{\chi}^2\right),
\label{Friedmann-Eq-1}\\
&& \ddot\phi+3\frac{\dot a}{a}\dot\phi+V'p_{\chi}^2=0,
\label{Friedmann-Eq-2}\\
&&\left(\frac{\dot a}{a}\right)^2=\frac13\left(\frac12 \dot\phi^2+Vp_{\chi}^2\right)
\label{Friedmann-Eq-3}\,,
\eeqa
where $p_{\chi}$ is a constant. As shown in \ref{sec2B}, setting $p_{\chi}=1$ reproduces
the Einstein equations and the equation of motion of $\phi$ in a
flat FLRW universe.


\subsection{Conformal Killing Vectors}
\label{sec3B}

We look for constants of motion
in the field space with the metric $G_{AB}$ \eqref{effective-metric}.

One immediately finds that the vector field $\xi=\partial/\partial \chi$ is a Killing vector field since the metric components do not depend on $\chi$.
Moreover, in the special case
when the potential $V$ is constant,
the metric components do not depend on $\phi$ and
additional Killing vector field
$\partial/\partial \phi$ arises.

We are interested in whether other conformal Killing vector
fields $\xi_A$ exist for some potential $V(\phi)$ so
that $\xi^Ap_A$ is a constant of motion.
This is accomplished by solving
the conformal Killing equations $\nabla_{(A}\xi_{B)}=fG_{AB}$.

In particular, we find that for the potential given by
\begin{align}
V(\phi) =
V_0 \left(
c_1 e^{\alpha \phi}
+c_2 e^{-\alpha \phi}\right)^{-2+\frac{\sqrt{6}}{\alpha}}
\label{potentila:conformal-k}
\end{align}
where $V_0,c_1,c_2$ and $\alpha$ are constants,
there exists a conformal Killing vector field $\xi_{(1)}$:
\begin{align}
\xi_{(1)} = -a^{-\sqrt{6}\alpha +1}
\left( c_1 e^{\alpha\phi}
- c_2 e^{-\alpha \phi} \right)
\frac{\partial}{\partial a}
+ \sqrt{6}a^{-\sqrt{6}\alpha}
\left(c_1 e^{\alpha\phi}
+ c_2 e^{-\alpha \phi}\right)
\frac{\partial}{\partial \phi} \,.
\label{CKV1}
\end{align}
$\xi_{(1)}$ satisfies the conformal Killing equations
$ \nabla_{(A}\xi_{(1)B)} = f G_{AB}$
with
\begin{align}
f =\frac13\nabla_A\xi^A_{(1)}= \sqrt{6}\left(\alpha-\frac{\sqrt{6}}{4}\right)
a^{-\sqrt{6}\alpha}\left( c_1 e^{\alpha\phi}
- c_2 e^{-\alpha \phi} \right) \,.
\end{align}
The derivation of the solution is outlined in Appendix \ref{appendixA}.
$\xi_{(1)}$ and $\xi_{(2)}=\p/\p\chi$ commute, $[\xi_{(1)}, \xi_{(2)}]=0$ and the Poisson bracket of two constants of motion is equal to zero, $\{\xi_{(1)}^Ap_A,\xi_{(2)}^Bp_B\}=0$.
Therefore, together with the Hamiltonian constraint Eq. (\ref{H-constraint}),
we have three functionally independent constants of motion for the system with three degrees of freedom, and
the system is completely integrable (in the sense of Liouville)
\footnote{According to Liouville's theorem, a Hamiltonian system with $n$ degrees of freedom is said to be completely integrable if there exist $n$ functionally independent constants of motion which are in involution (i.e. their Possion bracket is equal to zero). The system is then integrable by quadratures \cite{arnold}. }. This is the main result of this paper
\footnote{We note that there exists yet another conformal Killing vector field
$\xi_{(3)}=a\p/\p a+3\chi\p/\p\chi$ irrespective of the form of $V(\phi)$,
which is required for the integrability of the conformal Killing vector fields \cite{CH-inprep}. Therefore, for general $V(\phi)$ the equations of motion
(null geodesics on lifted field space) are reduced to the system of first-order equations for
$a,\phi$ and $\chi$.
However, $\xi_{(3)}$ does not commute with $\xi_{(1)}$ or $\xi_{(2)}$,
and the presence of the constant of motion
$\xi_{(3)}^Ap_A$ does not help to integrate the equations of motion by quadratures. }.

For $c_1=0$ or $c_2=0$ $V(\phi)$ (\ref{potentila:conformal-k}) becomes
\beqa
V(\phi)=V_0\exp\left[\pm\left(\sqrt{6}-2\alpha\right)\phi\right]\,,
\eeqa
which includes general exponential potential for power-law inflation
since $\alpha$ is an arbitrary constant.
\footnote{It is also interesting to note that in this case, the Cotton tensor vanishes and the field space is conformally flat.}

Note also that if $\alpha=\frac{\sqrt{6}}{4}$, then $f=0$ and
the conformal Killing vector field
becomes the Killing vector field found by \cite{deRitis:1990ba}.
Then, by introducing Killing coordinates along the Killing vector fields,
one may solve the equations of motion easily. Details may be found in
\cite{CH-inprep}.
Moreover,
the potential Eq. (\ref{potentila:conformal-k}) reduces to
\begin{align}
V(\phi) = V_0
\left(c_1 e^{\frac{\sqrt{6}}{4}\phi}
+ c_2 e^{-\frac{\sqrt{6}}{4}\phi}\right)^2 \,,
\label{potential-I}
\end{align}
which coincides with
that found by \cite{deRitis:1990ba} where the integrable
cosmological models are studied by searching for the Noether symmetry
for the Lagrangian.


\subsection{Construction of the Solutions for Power-law Inflation}


Given three constants of motion, $H, C=\xi^A_{(1)}p_A$ and $p_{\chi}$,
we can construct the solutions of the equations of motion by following
the proof of Liouville's theorem given in \cite{oy}. 
As an example, we consider the case with $c_1=0$ and $c_2=1$ so that
\beqa
V(\phi)=V_0e^{-\left(\sqrt{6}-2\alpha\right)\phi}\equiv V_0e^{-\lambda\phi}\,,
\eeqa
By setting $p_{\chi}=1$ we reduce the degrees of freedom, and
the two constants of motion are
\beqa
C&=&\xi_{(1)}^Ap_A=a^{-\sqrt{6}\alpha +1} e^{-\alpha \phi} p_a+
\sqrt{6}a^{-\sqrt{6}\alpha}
e^{-\alpha \phi}p_{\phi}\label{C}\,,\\
H&=& \frac{1}{2}\left(
-\frac{p_a^2}{6a}+\frac{p_\phi^2}{a^3}
+2a^3V(\phi) \right)=0\label{H}\,.
\eeqa
These equations are solved for $p_a$ and $p_{\phi}$
\beqa
p_a&=&\frac{C}{2}a^{\sqrt{6}\alpha-1}e^{\alpha\phi}+\frac{6V_0}{C}a^{5-\sqrt{6}\alpha}e^{(\alpha-\sqrt{6})\phi}\equiv f_a(a,\phi,C)\,,\\
p_{\phi}&=&\frac{C}{2\sqrt{6}}a^{\sqrt{6}\alpha}e^{\alpha\phi}-\frac{\sqrt{6}V_0}{C}a^{6-\sqrt{6}\alpha}e^{(\alpha-\sqrt{6})\phi}\equiv f_{\phi}(a,\phi,C)\,.
\eeqa

If we can introduce new canonical coordinates and momenta
$(Q_1,Q_2,P_1,P_2)$ by a canonical transformation such
that $P_1=C$ and $P_2=H$, then the canonical equations of motion become
$\dot Q_1=\p H/\p P_1=0, \dot Q_2=\p H/\p P_2=1$ and the solutions are trivially given by $Q_1={\rm const.}$ and $Q_2=t+{\rm const.}$ which
involve two additional constants.

Such a canonical transformation is provided by the following
generating function
\beqa
S(a,\phi,C)=\int f_a(a,\phi,C)da+f_{\phi}(a,\phi,C)d\phi\,.
\eeqa
Then, the new coordinates are given by ($i=1,2$)
\beqa
Q_i=\frac{\p S}{\p P_i}=\int\left(\frac{\p f_a}{\p P_i}da+
\frac{\p f_{\phi}}{\p P_i}d\phi \right)\,.
\eeqa
Moreover, from the definition of the new momenta, $C=P_1$ and $H=P_2$,
we have the following relations
\beqa
\begin{pmatrix}
\frac{\p C}{\p p_a} & \frac{\p C}{\p p_{\phi}} \\
\frac{\p H}{\p p_a} &\frac{\p H}{\p p_{\phi}} & \\
\end{pmatrix}
\begin{pmatrix}
\frac{\p f_a}{\p P_1}& \frac{\p f_a}{\p P_2}\\
\frac{\p f_{\phi}}{\p P_1}&\frac{\p f_{\phi}}{\p P_2}\\
\end{pmatrix}
=
\begin{pmatrix}
1&0\\
0&1\\
\end{pmatrix}
\,,
\eeqa
and $\frac{\p f_a}{\p P_i}$ and $\frac{\p f_{\phi}}{\p P_i}$ in the integrand can be expressed as functions of $a,\phi,p_a$ and $p_{\phi}$. Concretely $Q_1$ and $Q_2$ are written as
\beqa
Q_1&=&\int \frac{1}{\frac{\p C}{\p p_a}\frac{\p H}{\p p_{\phi}}-
\frac{\p C}{\p p_{\phi}}\frac{\p H}{\p p_a}
}\left(\frac{\p H}{\p p_{\phi}}da-\frac{\p H}{\p p_a}d\phi\right) \\
Q_2&=&\int\frac{1}{\frac{\p C}{\p p_a}\frac{\p H}{\p p_{\phi}}-
\frac{\p C}{\p p_{\phi}}\frac{\p H}{\p p_a}
}\left(-\frac{\p C}{\p p_{\phi}}da+\frac{\p C}{\p p_a}d\phi\right)\,,
\eeqa
and using Eq. (\ref{C}) and Eq. (\ref{H}), we obtain
\beqa
Q_1&=&\frac{12V_0}{(\sqrt{6}\alpha-6)C^2}a^{6-\sqrt{6}\alpha}e^{(\alpha-\sqrt{6})\phi}+\frac
{1}{\sqrt{6}\alpha}a^{\sqrt{6}\alpha}e^{\alpha\phi}
={\rm const.}\,,\label{Q1}\\
Q_2&=&\frac{6}{\alpha(2\alpha-\sqrt{6})C}a^{3-\sqrt{6}\alpha}e^{-\alpha\phi}=
t+{\rm const.}\label{Q2}\,.
\eeqa
We seek a particular solution by setting constants zero so that $Q_1=0$ and $Q_2=t$.
Then, from Eq. (\ref{Q1}), $e^{\phi}\propto a^{\sqrt{6}-2\alpha}$ and
putting this into Eq. (\ref{Q2}), we find
\beqa
&&a\propto t^{\frac{2}{(\sqrt{6}-2\alpha)^2}}=t^{\frac{2}{\lambda^2}}\,,\\
&&\phi=\frac{2}{\sqrt{6}-2\alpha}\ln t+{\rm const.}=\frac{2}{\lambda}\ln t +{\rm const.}\,,
\eeqa
thus reproducing the well-known solutions for power-law inflation \cite{Lucchin:1984yf}.


\section{Summary}
\label{sec7}

In order to search for symmetries of the system of a scalar field in FLRW universe,
we applied the Eisenhart lift for scalar fields to a scalar field in a flat
FLRW universe and found
that a conformal Killing vector field exists for a particular combination of
exponential functions
(\ref{potentila:conformal-k}) which includes the potential for power-law
inflation model.

The existence of the conformal Killing vector implies the existence of an additional conserved quantity and explains
why the scalar field as well as the scale factor
can be solved exactly for an exponential potential in FLRW universe.

\section*{Acknowledgments}
This work is supported by JSPS Grant-in-Aid for Scientific Research Number
22K03640 (TC) and in part by Nihon University.


\appendix

\section{Solving the Conformal Killing Equations}
\label{appendixA}

In this appendix, we briefly describe how the conformal
Killing equations are solved.
For the field space metric Eq. (\ref{effective-metric}),
the components of conformal Killing equations are
\beqa
-a\xi_a-3\frac{V'}{V}\xi_{\phi}+6\p_{\phi}\xi_{\phi}+2a^2\p_a\xi_a&=&0,\label{ckv1}\\
-\frac{3}{a}\xi_{\phi}+\p_{\phi}\xi_a+\p_{a}\xi_{\phi}&=&0,\label{ckv2}\\
-5a\xi_a+3\frac{V'}{V}\xi_{\phi}+12\p_{\phi}\xi_{\phi}+a^2\p_a\xi_a&=&0,\label{ckv3}\\
\frac{3}{a}\xi_{\chi}+\p_a\xi_{\chi}&=&0,\label{ckv4}\\
\frac{V'}{V}\xi_{\chi}+\p_{\phi}\xi_{\chi}&=&0,\label{ckv5}\\
4a\xi_a-6\frac{V'}{V}\xi_{\phi}-6\p_{\phi}\xi_{\phi}+a^2\p_a\xi_a&=&0\label{ckv6}\,.
\eeqa
Among these equations, Eq. (\ref{ckv4}) and Eq. (\ref{ckv5}) are solved by
the Killing vector $\xi_{(2)}=\p/\p\chi$.

We assume the factorized form of
$\xi_a=a^{\beta}f(\phi)$ and $\xi_{\phi}=a^{\gamma}g(\phi)$ with $\beta$ and $\gamma$ being constants. Then, from Eq. (\ref{ckv1}), $\gamma=\beta+1$. Also from Eq. (\ref{ckv2}),
$g=\frac{1}{2-\beta}f'$. Hence from Eq. (\ref{ckv1}) and Eq. (\ref{ckv3}), we obtain
\beqa
f''=\frac{(2-\beta)^2}{6}f\equiv \alpha^2f,
\eeqa
where $\alpha=(2-\beta)/\sqrt{6}$, and the solutions are given by
\beqa
f(\phi)&=&C_1e^{\alpha\phi}+C_2e^{\alpha\phi},\\
g(\phi)&=&\frac{f'}{2-\beta}=\frac{1}{\sqrt{6}}(C_1e^{\alpha\phi}-C_2e^{\alpha\phi})\,,
\eeqa
where $C_1 $ and $C_2$ are constants. 
Putting these into Eq. (\ref{ckv1}) gives $\frac{V'}{V}=\left(-2+\frac{\sqrt{6}}{\alpha}\right)\frac{g'}{g}$, which can be solved for $V(\phi)$ :
\beqa
V(\phi)\propto(C_1e^{\alpha\phi}-C_2e^{\alpha\phi})^{-2+\frac{\sqrt{6}}{\alpha}}\,.
\eeqa
We confirm that these solutions also satisfy Eq. (\ref{ckv6}).
Setting $C_1=6c_1$ and $C_2=-6c_2$ corresponds to the solution Eq. (\ref{CKV1}).

\bibliographystyle{apsrev4-1}
\bibliography{references}

\end{document}